\documentclass[%
footinbib,
superscriptaddress,
twocolumn,
amsmath,amssymb,
aps,
prl
]{revtex4-2}

\usepackage{ bbold }
\usepackage{graphicx}
\usepackage{dcolumn}
\usepackage{bm}
\usepackage[justification=RaggedRight]{caption}
\usepackage{subcaption}
\usepackage[utf8]{inputenc}
\usepackage[T1]{fontenc}
\usepackage{braket}
\usepackage{soul} 

\usepackage{dsfont}

\usepackage{xcolor}		
\usepackage[english]{babel} 

\newbox{\bigpicturebox}
\DeclareMathOperator\erf{erf}

\usepackage{csquotes}
\usepackage{hyperref}
\usepackage{bookmark}

\hypersetup{%
  breaklinks=true,
  colorlinks=true,
  linkcolor=black,
  filecolor=black,
  urlcolor=black,
  citecolor=black,
}

\begin{document}

\title{Dynamical systems on ultra small-world networks}

\author{Nirbhay Patil}

\author{Ada Altieri}
\affiliation{Laboratoire Matière et Systèmes Complexes (MSC), Université Paris Cité, CNRS, 75013 Paris, France}

\author{Fabi\'an Aguirre-L\'opez}
\affiliation{Université Paris 1 Panthéon-Sorbonne, CNRS, Centre d’Économie de la Sorbonne, Paris, France}

\begin{abstract}
Despite the knowledge that social, economical, and ecological networks are often of a small-world nature with inter-nodal distance growing even slower than logarithmically with system size, we often assume theoretical systems to be outside of this regime, to make them easier to treat analytically. Here we derive a framework to apply the powerful dynamical mean-field theory on highly heterogeneous networks that is able to account for more of the degree correlations naturally arising from network constraints, known as structural cut-offs. We apply this framework to the well-studied and understood disordered Lotka-Volterra model, and show typically reported observables such as survival rates and stability for these systems on ultra small-world networks. We find much better agreement for these variables for all ranges of exponents for simulated power-law networks as well as empirically sourced networks.
\end{abstract}

\maketitle

Statistical physics methods have long enjoyed a significant applicability to large complex models \cite{May1972,May1976,bouchaud2000wealth,Cu-Ku1993}, with special niches forming in the fields of theoretical ecology, epidemics, economic modelling, and more \cite{galla2018dynamically,Biroli2018_eco,Altieri-Biroli2021}. As our repertoire of the former grows over the years, so does our understanding of the latter \cite{poley2023generalized,garcia2024interactions,garnier2024unlearnable,patil2025emergent}. While many of these models assume systems with regular networks - often fully connected - empirical studies show that real networks display a variety of structures \cite{guimaraes2020structure,dunne2004network}. There has been a lot of work on the mean-field behaviour of complex systems over heterogenous networks, especially in the field of epidemic modelling \cite{dorogovtsev2008critical,pastor2001epidemic,pastor2015epidemic} and disordered systems \cite{metz2022mean,silva2022analytic,ferreira2023nonequilibrium,metz2025dynamical}. Recent works combine this with Dynamical Mean-Field Theory - a powerful tool that is able to represent the dynamics of a complex system in terms of the statistics of a single component in a self-consistent manner \cite{sompolinsky1981dynamic,crisanti1993spherical}. These studies show the behaviour of ecological models exhibiting non-trivial degree distributions \cite{park2024incorporating,poley2025interaction,aguirre2024heterogeneous}, developing the so called Heterogenous Dynamical Mean-Field Theory (HDMFT), with expressions for the stability of such systems, species abundance distributions, survival rates, and more.

In this work we aim to overcome one important limitation of the current HDMFT theory, which is its assumption of small degree heterogeneities in the distributions taken. We deem high heterogeneities to represent a crucial regime of ecological systems, as evidenced in \cite{montoya2002small,jordano2003invariant}. These studies analyse a large number of ecosystems from previous works, showing that a majority of them are power-law or truncated power-law distributed networks \cite{mossa2002truncation}, with a degree distribution of the form
\begin{align}
\label{eq:pareto-bounded}
    P(k)\propto k^{-1-\alpha}.
\end{align}
The values of the exponent $\alpha$ cited in the papers are extremely small, ranging from 0.07 to 1.78 \cite{jordano2003invariant,wheelwright1984tropical,tutin1997primate}  - values for which we would observe effects of a so called structural cut-off due to the size of the system, with typical connectivities attaining values comparable to the total number of nodes.

\begin{figure}
    \centering
    \includegraphics[height=4.2cm]{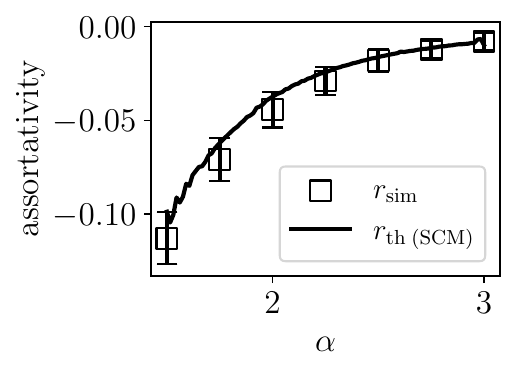}
    \caption{The constraint that any node can have no more than $N$ connections causes highly heterogeneous networks to have non-trivial correlations, with a non-zero assortativity coefficients $r$ without being constructed specifically to be assortative. Here we show that networks simulated for varying values of the pareto exponent $\alpha$ with fixed $N=4000$ and $C=30$ are forcibly negatively assorted. The theoretically expected value of the assortativity is 0 if calculated using the statistically independent edge probability function $P_{\rm lin}=km/CN$, but including the effect of the hidden degrees  $P_{\rm SCM}(\tilde k(k),\tilde m(m))$ we are able to theoretically take into consideration this assortativity.}
    \label{fig:assort}
\end{figure}

This high heterogeneity property poses one major issue for the applicability of the HMDFT  introduced for the generalized Lotka-Volterra model  \cite{park2024incorporating,poley2025interaction,aguirre2024heterogeneous}, since it assumes an underlying maximum entropy random network with a degree distribution with properly bounded moments of its degree distribution. In particular, the annealed approximation presented there assumes that all networks come from the Chung-Lu model, \cite{chung2002connected}, where edges $(i,j)$ are statistically independent and exist with probability
\begin{align}
\label{eq:chung-lu}
    p_{ij} = \frac{k_i k_j}{NC}.
\end{align}between nodes with degree $k_i$ and $k_j$ respectively, for a network with $N$ nodes
 with $C$ being the average number of edges per node. This implies that the degrees in the network satisfy the condition
\begin{align}
    k_i k_j \le  NC,
\end{align}
which can be easily violated by networks with a power law degree distribution like in eq.\eqref{eq:pareto-bounded}.

We overcome this issue by mapping the observed degrees of a network to a range of ``hidden degrees'' that are considered to be secretly forming the backbone of the network. This is to say that the probability $P(\tilde k,\tilde m)$ of two nodes being connected is a function of certain unknown variables $\tilde k$ and $\tilde m$ corresponding to them, written as \begin{align}
    \begin{aligned}
        P_{\rm SCM}(\tilde k,\tilde m)&=\frac{\tilde k\tilde m}{\tilde k\tilde m+CN}.
    \end{aligned}\label{eq:PSCM}
\end{align}The above expression is obtained from the maximum entropy ensemble of networks with a provided degree distribution, also known as the soft configurational model. The maximum entropy ensemble gives the set of networks under a chosen constrain - providing graphs chosen as randomly as possible within the range of permissible networks, resulting in as little biases or extra correlations as possible. The unknown hidden degree variables are equivalent to a rescaling of the Lagrange multipliers or the fugacity \cite{park2004statistical} of the network used to fix the degree under the maximum entropy framework. Given a set of hidden degrees, we can write them as a continuous variable with density $\tilde \rho(\tilde m)$ for large networks. The true degree $k$ of a node with hidden degree $\tilde k$ is then
\begin{align}
    k(\tilde k)&=N\int d\tilde m P(\tilde k,\tilde m)\tilde \rho(\tilde m).\label{eq:kasftildk}
\end{align}
One can now invert eq.(\ref{eq:kasftildk}) numerically given any observed network to then deduce the distribution of the hidden degrees.  If instead the observed network is known to be power-law distributed, the true degree can be expressed analytically to be
\begin{align}
    \begin{aligned}
        k=N {}_2F_1\left(1,\alpha,1+\alpha;-\frac{ CN}{\tilde k_0\tilde k}\right),
    \end{aligned}\label{eq:hypgeom}
\end{align}
where $\tilde k_0$ is the lower cut-off of the hidden degrees chosen such that the observed degrees have the correct average \cite{park2003origin}. This can be then inserted in the expression for the joint degree probabilities which are known as a function of the hidden degrees to write $P(\tilde k(k),\tilde m(m))$, where $\tilde k(k)$ is given by the inverse of the relations in eq.(\ref{eq:kasftildk}) or (\ref{eq:hypgeom}). For small values of the hidden degrees $\tilde k\ll\sqrt{CN}$ the observed degrees are found to match them exactly, in which limit one can write the joint probability as $P(k,m)=km/CN$, the bilinear function used previously in the development of HDMFT  \cite{park2024incorporating,aguirre2024heterogeneous,poley2025interaction}. We use the term ``hidden degree'' to highlight this relation and similarity of the hidden and observed degrees in this limit of small connectivity. Outside of this limit, the choice to use the full expressions avoids disregarding the important degree correlations that are bound to happen in the presence of nodes with very high connectivity. This can be seen by the non-trivial assortativity present in these networks that can be explained solely by the degree distribution, as seen in fig.(\ref{fig:assort}).  The Chung-Lu model is unable to account for these degree correlations since it assumes degrees are uncorrelated by definition. This disassortative nature of the graphs is born out of structural cut-offs, which is unavoidable in scale-free networks of finite size \cite{boguna2004cut,catanzaro2005generation,zhou2007structural}. These structural effects have been previously taken into account in epidemic modelling for systems without edge disorder, \cite{boguna2002epidemic}. In the following we will show how the effect of disorder can be included.

To demonstrate that we can make theory that is both mean-field type and that takes into account the correlations, we shall study the well explored Generalised Lotka-Volterra (GLV) model, with the growth rate of the population of a species being given by
\begin{align}
    \begin{aligned}
        \dot S_i(t)=S_i(t)\left(1-S_i(t)-\sum_jA_{ij}\beta_{ij}S_j(t)\right),
    \end{aligned}
    \label{eq:model}
\end{align}
where $S_i(t)$ is the abundance of species $i$ at time $t$. The object $\beta_{ij}$ depicts the strength of the interaction between species $j$ and $i$ on the species $i$, while $A_{ij}$ is the adjacency matrix, its elements being 1 or 0 based on whether or not this interaction exists. We assume that the adjacency matrix is symmetric, i.e. if species $i$ has an effect on species $j$, the opposite is likely true as well. The interaction strengths on the other hand may or may not have correlations. Both the interaction and the adjacency matrices are assumed to be sources of disorder, with the interactions strengths satisfying
\begin{align}
    {\rm mean}[\beta_{ij}]=\frac{\mu}{C},\;\;\; {\rm var}[\beta_{ij}]=\frac{\sigma^2}{C},\;\;\; {\rm corr}[\beta_{ij}\beta_{ji}]=\frac{\gamma\sigma^2}{C}.
\end{align}
Here we would have $\mu>0$ for a system that has more competitive interactions, and $\mu<0$ for one that is more cooperative. The standard deviation $\sigma$ captures the strength or randomness of the disorder, and the correlation $\gamma$ incorporates whether the interactions tend to be reciprocal. We work with zero correlations $\gamma=0$ for the rest of the paper to showcase expressions that can be followed more easily, but the full expressions with non-zero reciprocity are also presented in the appendix. The ensemble of adjacency matrices that we use simply need to have a prescribed degree distribution. As the elements of these matrices are simply 0 or 1 with probability $p_{ij}$, their statistics are given by
\begin{align}
    \begin{aligned}
        \langle A_{ij}^n\rangle=p_{ij},
    \end{aligned}
\end{align}
for any value of $n>0$.

\begin{figure}
    \centering
    \includegraphics[width=\linewidth]{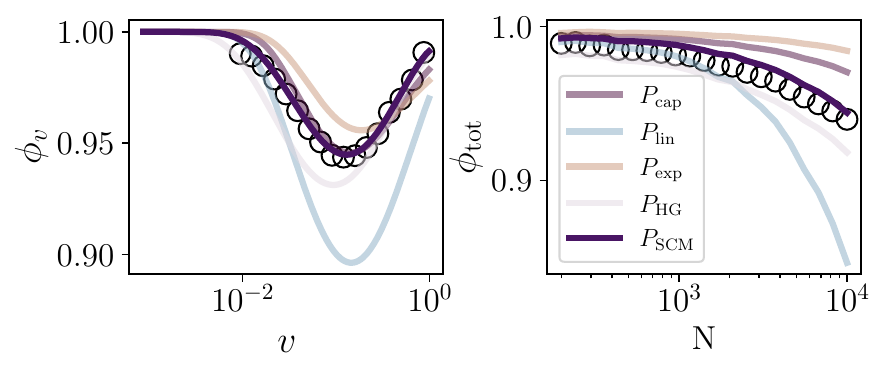}
    \caption{We compare the predicted survival rate for the various assumptions of the joint distribution function $P(k,m)$ with $\alpha=1.5$, $\mu=-0.1$, and $\sigma=0.3$. On the left we plot the survival rate as a function of connectance $v=k/N$ for $C=26$,  and $N=1162$. On the right we show the total survival rate as a function of the system size, while varying $C\propto \sqrt N$. We find the differences between the predictions of these assumptions to vary over a larger range in the regime with mutualistic interactions. The theoretical predictions given by the joint degree distribution functions that factor in the difference between hidden and observed degrees match the simulation results best.}
    \label{fig:phiv_andNconvergence}
\end{figure}

\begin{figure}
    \centering
    \includegraphics[width=\linewidth]{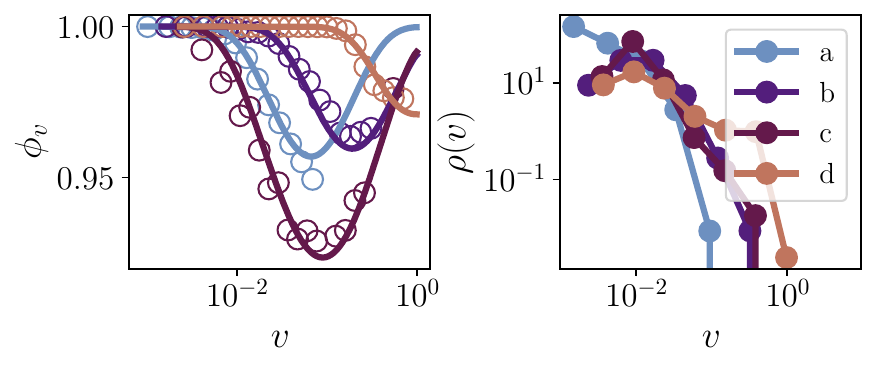}
    \caption{Survival probabilities (left) as a function of connectivity $v$ for $\mu=-0.1$, $\sigma=0.3$ for different networks obtained from \cite{networksnetz}. The distributions of the connectances $\rho(v)$ are given on right, and correspond to data from \cite{guimera2003self,dunne2014highly,duch2005community,slattery1998combining} in order from (a) to (d). Two of these were asymmetric/directed networks, for which we retained half the edges and made them undirected. The four networks have sizes $N_{(a)}=1133$, $N_{(b)}=700$, $N_{(c)}=453$, and $N_{(d)}=434$, with average degree $C_{(a)}=9.6$, $C_{(b)}=18.3$, $C_{(c)}=9$, and $C_{(d)}=70.2$. A power-law fit for each would give exponents $\alpha_{(a)}=1.18$, $\alpha_{(b)}=0.46$, $\alpha_{(c)}=0.66$, and $\alpha_{(d)}=0.24$.}
    \label{fig:realnets}
\end{figure}

Using generating functional analysis we derive the following Heterogeneous Dynamical Mean-Field Theory, 
\begin{align}
    \dot S_k(t)=S_k(t)\left(1-S_k(t)-\mu B_k(t)+\eta_k(t)\right),
\label{eq:HDMFT}
\end{align}
which represents the effective dynamics for one species with degree $k$. This framework approximates the effects of a large number of random variables by their statistics, requiring the assumption that the number of edges for most nodes is large enough. Here the noise $\eta_k(t)$ has zero mean, and correlation $\langle\eta_k(t)\eta_k(t')\rangle=q_k(t,t')$. The parameters $B_k(t)$ and $q_k(t,t')$  are functions of the degree as well, having the relations
\begin{align}
\label{eq:self-consistent-equations}
    \begin{aligned}
        cB_k(t)&=\int dm \rho(m) P_{km} \Big\langle S_m(t)\Big\rangle,\\
                cq_k(t,t')&=\int du \rho(m)P_{km}\Big\langle S_m(t)S_m(t')\Big\rangle,
    \end{aligned}
\end{align}
where $\rho(m)$ is the distribution of degrees, $P_{km}$ is the probability of nodes with degree $k$ and $m$ being connected, and $c=C/N$. The average is over the coloured noise distributions and initial conditions. Once these functions, $B_k(t)$ and $q_k(t,t')$ are determined self-consistently from eq.\eqref{eq:self-consistent-equations}, then one can calculate the expected value of population averages in the following way,
\begin{align}
    \left\langle \frac{1}{N} \sum_{i=1}^N f(S_i(t))\right\rangle = \int dm\rho(m) \langle f(S_m(t))\rangle 
\end{align}
where we are averaging over $A_{ij}$, $\beta_{ij}$, and initial conditions on the LHS and over the effective stochastic process \eqref{eq:HDMFT} on the RHS. Notice this theory is considerably more complex than other previous approaches to HDMFT since we need to calculate two moments, $B_k(t)$ and $q_k(t,t')$ \emph{per degree} and these are coupled through eq.\eqref{eq:self-consistent-equations}. Nevertheless, even though all the moments are coupled, averages over the noise \emph{are still independent}. When there exists a unique fixed point for eq.\eqref{eq:model}, these parameters reach a constant value, and the species populations can be written as
\begin{align}
    \begin{aligned}
        S^*_k={\rm max}(0,1-\mu B_k^*+\sigma\sqrt{q_k^*}z),
    \end{aligned}
\end{align} with  $z$ being a gaussian random variable. In this case we can appreaciate the reduction in complexity given by the theory. The species survival rate is then written in a similar form as for the fully connected system but depends on the connectivity of the species in question and the corresponding statistical parameters at equilibrium, with
\begin{align}
    \phi_k=\frac{1}{2}\left[1+\erf\left(\frac{1-\mu B_k^*}{\sigma\sqrt{2q_k^*}}\right)\right].
\end{align}
The total species survival can be calculated by integrating this over the degree distribution. 
In fig.(\ref{fig:phiv_andNconvergence}) we plot the survival rate of a species as a function of its connectivity as observed in simulations, and the prediction obtained for the same with five different assumptions of the joint degree probabilities. The bilinear function $P_{\rm lin}(k,m)=km/CN$ is what has been used in theory so far \cite{park2024incorporating,aguirre2024heterogeneous,poley2025interaction}. For extremely heavy tailed distributions you can have multiple hubs with large degree, this would result in an unphysical probability of connection $P(k,m)>1$. This overestimation of connection probability has been explained by the fact that in these kinds of models, two nodes cannot share more than one edge \cite{maslov2004detection}. The simplest solution would be to set $P_{\rm cap}(k,m)={\rm min}(km/CN,1)$ \cite{serrano2026statistical}. One can also instead take the probability of connection coming from various algorithms of network creation, like the static model \cite{goh2001universal} which gives $P_{\rm exp}(k,m)=1-\exp(-km/CN)$, where links are created between nodes picked by weights proportional to the degree. Finally, we consider the maximum entropy model corresponding to a network ensemble with the observed degrees, resulting in the correlations given in eq.(\ref{eq:PSCM}). Inserting the numerical inverse of eq.(\ref{eq:kasftildk}) or the inverse of eq.(\ref{eq:hypgeom}) for the hidden degrees in this correlation function gives the lines corresponding to $P_{\rm SCM}$ and $P_{\rm HG}$ approximations respectively. In fig.(\ref{fig:realnets}), we show the survival rates of Lotka-Volterra simulations running on various networks sourced from real data \cite{guimera2003self,dunne2014highly,duch2005community,slattery1998combining}, compared to the theoretical prediction given by the HDMFT that uses the edge probability $P_{\rm SCM}$, back-calculating the hidden degrees of these networks numerically using as few as 10 bins for the degree distribution.

\begin{figure}
    \centering
    \includegraphics[height=3.2cm]{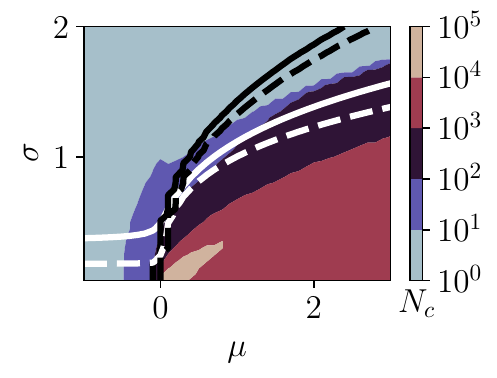}
    \includegraphics[height=3.2cm]{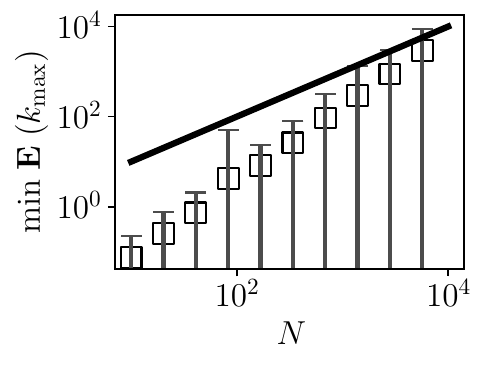}
    \caption{On the left we plot the value of the system size $N_c$ beyond which any stable system would necessarily have degrees of the same order as $N_c$ for a power-law pareto exponent $\alpha=1.5$. This highlights the importance of degree correlations in different regimes of the phase space for different values of the system size. The dashed lines are the instability predictions to multiple equilibria (white) and unbounded growth (black) without considering degree correlations, while the solid lines are the theoretical predictions including degree correlations. On the right we show for one set of values $\mu=-0.1$, $\sigma=0.3$ and $\alpha=1.5$, the maximum degree expected from a system where we set $C$ to be the minimum accepted value below which the system would be unstable, $C=c_{\rm min}N$. The solid black line marks the parity line $y=x$.}
    \label{fig:Ncrit}
\end{figure}

The linear stability condition for the single equilibrium solution can be expressed as
\begin{align}
    \begin{aligned}
        {\rm Max}\left[{\rm Eig}\left(\frac{\sigma^2}{c}\rho_m\phi_kP_{km}\right)\right]<1,
    \end{aligned}\label{eq:UFPstab}
\end{align}
requiring the operator within the parentheses to have no eigenvalues larger than 1. Notice that the relations in \eqref{eq:self-consistent-equations} and \eqref{eq:UFPstab} all simplify to the previously studied version of HDMFT \cite{aguirre2024heterogeneous,park2024incorporating,poley2025interaction} if the degree correlations are not present, that is, if we take $P_{km}=P_{\rm lin}=km/CN$.

We find that the system gets more unstable as $c$ decreases, such that for a given set  of parameters $\mu$, $\sigma$, and exponent $\alpha$, there is a critical value $c_{\rm min}=f(\mu,\sigma,\alpha)$, below which the system would be unstable. This implies that the maximum expected degree of a graph - given by $k_{\rm max}\sim CN^{1/\alpha}=f(\mu,\sigma,\alpha)N^{1+1/\alpha}$ - has a minimum value for any stable system. This implies that for large stable systems, there are necessarily large degrees of the order of the size of the system. We show this critical system size as a function of the disorder statistics in fig.(\ref{fig:Ncrit}). Here the solid white line shows the phase boundary between unique and multiple fixed points by checking the linear stability condition of the unique fixed point \eqref{eq:UFPstab} using $P_{\rm SCM}$, while the dashed white line is the same condition checked within the annealed approximation $P_{\rm lin}$. The black lines similarly show the boundary of a phase with unbounded growth, calculated by checking numerically for divergence of the average abundance or equivalently $B_k$. We see that for the most stable parts of the phase diagram with low $\mu$ and $\sigma$, one could possibly use the bilinear simplification $P_{\rm lin}$ without running into unaccounted correlations up till a system size of $N=10^5$, but as we get closer to the phase boundaries, any stable system bigger than $N=10^3$ would have hubs comparable to the size of the system, making it necessary to consider the non-zero assortativity. 

\section{Discussion}

Highly heterogeneous networks end up having correlations and non-zero assortativity coefficients due to the structural cut-off constraints applicable on a real network. In this work we show how these correlations can be factored into the edge probability used to derive the HDMFT equations by accounting for the relation between the hidden and true degrees using a maximum entropy ensemble of networks with fixed degree distributions. Using the survival rates of species in a Lotka-Volterra model as an important practical observable that can be used as a measure of the accuracy of our predictions, we show that including these correlations lets us predict the behaviour of extremely varied networks, including artificially simulated ones that resemble power laws as well as ones coming from citation networks, metabolic networks, food webs, and university email networks.  It has been shown before that real life networks found in biological and social systems tend to be disassortative \cite{newman2002assortative,watts1998collective,jeong2001lethality}, which seems to naturally be the case for the extremely heterogeneous networks that we study. Disassortative networks are interesting for a number of reasons - they are more prone to spread of features such as disease, and also more vulnerable to being broken by deletion of important nodes. Here we have developed a DFMT-type framework that can actually account for this naturally present artifact of many real networks. In the case of the Lotka-Volterra model, we find that this unaccounted for disassortativity increases survival rates and stability, not unlike the effect of negative correlations between interaction coefficients.

Our results are able to account for sparse and dense networks along with correlations, provided the correlations are a result of organic constraints necessary for the existence of a given degree distribution with high connectivity. For positively assortative networks this may not be the case, making that a natural direction for future research. Directed and bipartite networks or systems with other forms of internal structure \cite{garlaschelli2003universal,garlaschelli2004patterns,patil2024spectral,baron2020dispersal,annibale2009tailored} are also an interesting avenue for further work. 

\section{Acknowledgements}

We would like to thank S. Azaele, T. Arnoux de Pierry, and J. Baron for valuable feedback on this work. AA and NP acknowledge the support of ANR JCJC ‘SIDECAR’ ANR-23-CE30-0012-01. FAL acknowledges the support of ANR-24-CPJ1-0114-01. 


\bibliography{Bibliog}

\appendix

\onecolumngrid

\section{Generating function approach to Dynamical Mean-Field calculations}

We use the typical generating function method with path integrals over the dynamical equations for all species $S_i$ where the adjacency matrix is given by its elements $A_{ij}$ that are $0$ or $1$, and the interaction coefficients are $\beta_{ij}=\frac{\mu}{C}+\frac{\sigma}{\sqrt{C}}\xi_{ij}$ where $C$ is the average degree of the nodes in the network. The interactions have correlations $\langle \xi_{ij} \xi_{ji}\rangle=\gamma$. The generating function is then
\begin{align}
\begin{aligned}
Z[\boldsymbol{\psi}]=\int & \mathcal{D}[\dots]\exp \Biggl \lbrace i \sum_i \int dt  \hat{S}_i(t) \left[ \frac{\dot{S}_i}{S_i}-\left(1- S_i-\sum_{j} A_{ij}\beta_{ij}S_j +\epsilon_i \right) \right] \Biggr \rbrace  \exp \left(i \sum_i \int dt \; S_i(t) \psi_i(t) \right)\\
 =\int &\mathcal D[\dots]\mathcal M\exp \left(i \sum_i \int dt \; S_i(t) \psi_i(t) \right) \times \exp\Biggr\lbrace- i\sum_i\int dt\left[\hat S_i(t)\sum_jA_{ij}\beta_{ij}S_j\right]\Biggr\rbrace
\end{aligned}
\end{align}
where
\begin{align}
    \begin{aligned}
        \mathcal M&=\exp\Biggl \lbrace i \sum_i  \int dt  \left[ \hat{S}_i(t)\left( \frac{\dot{S}_i}{S_i}-1 + S_i+\epsilon_i \right)  \right]\Biggr \rbrace 
    \end{aligned}
\end{align}
Now consider only the averages over $\beta$ and $A$
\begin{align}
    \begin{aligned}
        &\overline{\exp\Biggr\lbrace-i\sum_{i<j}\int dt\left(A_{ij}\hat S_i\beta_{ij}S_j+A_{ij}\hat S _j\beta_{ji}S_i\right)\Bigg\rbrace}\\
        =&\overline{\prod_{i<j}\left(1+P(A_{ij}=1)\left[\exp(\int dt(-i\hat S_i\beta_{ij}S_j-i\hat S_j\beta_{ji}S_i))-1\right]\right)}\\
        =&\overline{\prod_{i<j}\left(1+P(A_{ij}=1)\left[\exp\int dt\left(-i\frac\mu C\hat S_iS_j-i\frac\mu C\hat S_jS_i-\int dt'(\frac{\sigma^2}{2C}\hat S_i\hat S_i'S_jS_j'+\frac{\sigma^2}{2C}\hat S_j\hat S_j'S_iS_i'+\frac{\sigma^2\gamma}{C}\hat S_i\hat S_j'S_i'S_j)\right)-1\right]\right)}\\
        \approx&\overline{\prod_{i<j}\exp\Biggr\lbrace \int dtP(A_{ij}=1)\left(-i\frac\mu C\hat S_iS_j-i\frac\mu C\hat S_jS_i-\int dt'(\frac{\sigma^2}{2C}\hat S_i\hat S_i'S_jS_j'+\frac{\sigma^2}{2C}\hat S_j\hat S_j'S_iS_i'+\frac{\sigma^2\gamma}{C}\hat S_i\hat S_j'S_i'S_j)\right)\Biggr\rbrace}
    \end{aligned}
\end{align}
where we're hoping the average degree is big enough for mean-field arguments to be valid and higher order terms in $1/C$ won't be significant. Now let's say
\begin{align}
    \begin{aligned}
        P(A_{ij}=1)&=P(\tilde k_i\tilde k_j),\\
        &=\sum_{n}a^2_{n}(\tilde k_i\tilde k_j)^n,
    \end{aligned}
\end{align}where $\tilde k$ is the hidden degree that's a function of the true degree $k$ in a known way. This lets us define the order parameters
\begin{align}
    \begin{aligned}
        b_n(t)&=\frac1Na_n\sum_i\tilde k_i^n\hat S_i(t)&\qquad d_n(t)&=\frac1Na_n\sum_i\tilde k_i^n S_i(t)\\
        q_n(t,t')&=\frac1Na_n\sum_i\tilde k_i^n\hat S_i(t)\hat S_i(t')&\qquad r_n(t,t')&=\frac1Na_n\sum_i\tilde k_i^n S_i(t)S_i(t')\\
        &&X_n(t,t')=\frac1Na_n\sum_i\tilde k_i^n\hat S_i(t)S_i(t')&
    \end{aligned}
\end{align}
So the moment generating action is
\begin{align}
\begin{aligned}
\int &\mathcal D[\dots]\mathcal M\exp \left(i \sum_i \int dt \; S_i(t) \psi_i(t) \right) \exp\Biggr\lbrace-N\int dt\sum_n\left[\frac\mu cb_n(t)d_n(t)+\int dt'\left(\frac{\sigma^2}{2c}q_n(t,t')r_n(t,t')+\frac{\sigma^2\gamma}{c}X_n(t,t')X_n(t',t)\right)\right]\Biggr\rbrace\\
&\times \exp\Biggr\lbrace i\int dt\sum_n\left[\hat b_n\left(Nb_n-\sum_ia_n\tilde k_i^n\hat S_i\right)+\hat d_n\left(Nd_n-\sum_ia_n\tilde k_i^n S_i\right)\right]\Biggr\rbrace\\
&\times \exp\Biggr\lbrace i\int dt\int dt'\sum_n\left[\hat q_n\left(Nq_n-\sum_ia_n\tilde k_i^n\hat S_i\hat S_i'\right)+\hat r_n\left(Nr_n-\sum_ia_n\tilde k_i^n S_iS_i'\right)\right]\Biggr\rbrace\\
&\times \exp\Biggr\lbrace i\int dt\int dt'\sum_n\left[\hat X_n\left(NX_n-\sum_ia_n\tilde k_i^n\hat S_i  S_i'\right)\right]\Biggr\rbrace
\end{aligned}
\end{align}
where $c=C/N$. This gives $b_n=\hat d_n=q_n=\hat r_n=0$ on maximising with respect to the order parameters, and
\begin{align}
    \begin{aligned}
        \hat b_n(t)&=-\frac{\mu}{c}d_n(t)\\
        \hat q_n(t,t')&=-\frac{\sigma^2}{2c}r_n(t,t')\\
        \hat X_n(t,t')&=-\frac{\sigma^2\gamma}{c}X_n(t',t)
    \end{aligned}
\end{align}
The remaining terms in the moment are then
\begin{align}
    \begin{aligned}
        \int &\mathcal D[\dots]\mathcal M\exp \left(i \sum_i \int dt \; S_i(t) \psi_i(t) \right) \exp\Biggr\lbrace-N\int dt\sum_n\left[\frac \mu cd_na_n\tilde k_i^n\hat S_i+\int dt'(\frac{\sigma^2}{2c}r_na_n\tilde k_i^n\hat S_i\hat S_i'+\frac{\sigma^2\gamma}{c}X_na_n\tilde k_i^n\hat S_iS_i')\right]\Biggr\rbrace
    \end{aligned}
\end{align}
Now the three terms on the right can be rewritten in terms of the edge probability, by observing that
\begin{align}
    \begin{aligned}
        \sum_nd_na_n\tilde k_i\hat S_i&=\sum_n\langle a_n\tilde k^nS\rangle a_n\tilde k_i\hat S_i\\
        &=\hat S_m\langle \sum_na_n\tilde k^n\tilde m^n S_k\rangle_{\tilde k}\\
        &=\hat S_m\langle P(\tilde k\tilde m)S_k\rangle_{k}\\
        &=\hat S_mB_k,
    \end{aligned}
\end{align}
where the indices are changed to $k$ and $m$ to represent species with $k$ or $m$ degrees, and $B_k$ is a new order parameter similar to $b_n$ but a function of the degree instead of being part of an infinite polynomial sum. Performing such recombinations for all three terms tells us that the effective dynamics of a species with degree $k$ is
\begin{align}
    \begin{aligned}
        \dot S_k(t)=S_k(t)\left(1-S_k(t)-\mu B_k(t)-\eta_k(t)-\sigma^2\gamma\int dt'R_k(t,t')S_k(t')\right),
    \end{aligned}
\end{align}
where the order parameters to be determined self consistently are
\begin{align}
    \begin{aligned}
        cB_k(t)&=\Bigg\langle\int dm\rho(m)P(\tilde k(k)\tilde m(m))S_m(t)\Bigg\rangle\\
        cq_k(t,t')&=\Bigg\langle\int dm\rho(m)P(\tilde k(k)\tilde m(m))S_m(t)S_m(t')\Bigg\rangle\\
        cR_k(t,t')&=\Bigg\langle\int dm\rho(m)P(\tilde k(k)\tilde m(m))\frac{\delta S_m(t)}{\delta \eta_m(t')}\Bigg\rangle,
    \end{aligned}
\end{align}
where $q_k$ are the correlations of the coloured noise $\langle \eta_k(t)\eta_k(t')\rangle=\sigma^2q_k(t,t')$. In the case where $\gamma=0$, the equilibrium values of the above order parameters then are
\begin{align}
    \begin{aligned}
        cB_k^*&=\int dk\rho(k)P(\tilde k(k),\tilde m(m))\int dz(1-\mu B_m^*-\sigma\sqrt{q_m^*}z)\Theta(1-\mu B_m^*-\sigma\sqrt{q_m^*}z)\\
        &=\int dm\rho(m)P(\tilde k(k),\tilde m(m))\sigma\sqrt{q_m}\left[\frac{e^{-\Delta_m^2/2}}{\sqrt{2\pi}}+\frac{\Delta_m}{2}(1+{\rm erf}(\frac{\Delta_m}{\sqrt{2}}))\right],\\
        cq_k^*&=\int dm\rho(m)P(\tilde k(k),\tilde m(m))\int dz(1-\mu B_m^*-\sigma\sqrt{q_m^*}z)^2\Theta(1-\mu B_m^*-\sigma\sqrt{q_m^*}z)\\
        &=\int dm\rho(m)P(\tilde k(k),\tilde m(m))\sigma^2q_m\left[\frac{\Delta_me^{-\Delta_m^2/2}}{\sqrt{2\pi}}+\frac{1+\Delta_m^2}{2}(1+{\rm erf}(\frac{\Delta_m}{\sqrt{2}}))\right],\\
        {\rm with}\;\;\; \Delta_k&=\frac{1-\mu B_k}{\sigma \sqrt{q_k}}.
    \end{aligned}
\end{align}
The linear stability of the equilibrium solution can be analysed by perturbing it with a noise $\delta h$ and taking a Fourier transform,
\begin{align}
    \begin{aligned}
        \delta \dot S_k&=S_k^*\left(-\delta S_k+\delta \eta_k+\delta h\right)\\
        \left(\frac{i\omega}{S^*_k}+1\right)\delta \tilde S_k&=\delta\tilde \eta_k+\delta \tilde h\\
        \langle|\delta \tilde S_k|^2\rangle&=\frac{\langle |\delta \tilde \eta_k|^2\rangle+\langle |\delta  \tilde  h|^2\rangle}{\left(\frac{i\omega}{S_k^*}+1\right)^2}=\frac{\phi_k\sigma^2q_k+\langle |\delta  \tilde  h|^2\rangle}{\left(\frac{i\omega}{S_k^*}+1\right)^2}
    \end{aligned}
\end{align}
where we use the fact that the coloured noise has correlations given by the order parameter $q_k$, and that these equations are only applicable to the surviving fraction $\phi_k$ of the species. Sticking to the zeroth Fourier mode $\omega=0$, we can then write
\begin{align}
    \begin{aligned}
        &\int dm \left[\delta (k-m)-\frac{\sigma^2}{c}\phi(k)\rho(m)P(k,m)\right]\langle|\delta\tilde S_m|^2\rangle=\langle|\delta \tilde h|^2\rangle\\
        &\implies \int dm \left[\delta (k-m)-\frac{\sigma^2}{c}\phi(k)\rho(m)P(k,m)-\frac{\langle |\delta \tilde h|^2\rangle}{Z}\right]Y_m=0,
    \end{aligned}
\end{align}
where $Y_k=\frac{\langle |\delta \tilde S_k|^2\rangle}{Z}$, and $Z=\int dm\langle |\delta \tilde S_m|^2\rangle$ using the method of \cite{poley2025interaction}. Now in the linearly unstable regime, the perturbation diverges, meaning $Z\rightarrow\infty$. This lets us ignore the term with the $Z$ in the denominator, making this 
\begin{align}
    Y(k)=\int dm\frac{\sigma^2}{c}\phi(k)\rho(m)P(k,m)Y(m),
\end{align}
which is equivalent to an eigenvector equation for a vector $Y_k$, acted upon by the matrix that is the function in the integral above. Thus we have instability when this matrix has eigenvalues greater than or equal 1.

\section{Numerical methods and convergence}

For Lotka-Volterra simulations, we sample a list of degrees from a given distribution and create it using the stubs method. Interaction coefficients are then sampled from a normal distribution, and initial values for species populations are taken uniformly between 0 and 1. For the dynamics we use the Runge-Kutta-Fehlberg method, with any simulation considered to have converged if $|\Delta S_i|<5\times10^{-13}$ for all species. Any species with a population under $10^{-10}$ is considered to be extinct.
\begin{figure}
    \centering
    \includegraphics[width=0.25\linewidth]{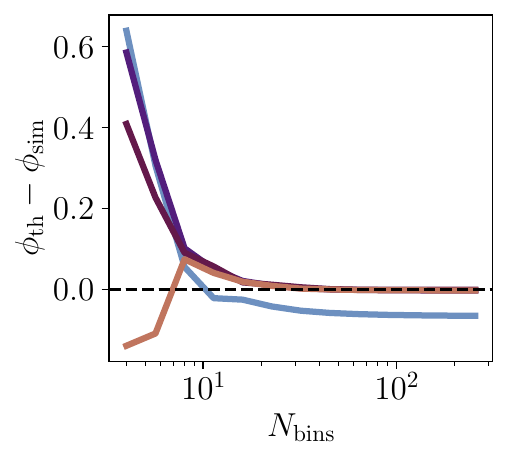}
    \caption{For real networks, the theoretical results converge quite quickly, giving the same results for any $N_{\rm bins}>30$. }
    \label{fig:binconv}
\end{figure}

The theoretical parameters are calculated self-consistently using iterative methods, representing each continuous functional parameter as an array that iteratively approaches the function that satisfies the conditions required. The hidden degrees are also calculated in a similar manner. Theoretically for a system with $N$ species each one would have to solve for $N$ hidden degrees along with the continuous functions $B_k$ and $q_k$ that would have almost as many degrees of freedom, but we reduce this complexity by making the mapping between hidden and observed degrees to be one between two continuous functions rather than between a precise sets of values. This lets us take whatever granularity we wish to work with, and we find that using more than 30 bins is redundant (fig.(\ref{fig:binconv})). We perform this numerical inversion by dividing the distribution into a number of bins, and then iterating over the integral
\begin{align}
    \tilde k_p^{(n+1)}=\frac{k_p}{N\sum_r d\tilde k^{(0)}_r\tilde \rho^{(0)}_r(\tilde k^{(0)}_r)\frac{\tilde k^{(n)}_r}{CN+\tilde k^{(n)}_p\tilde k^{(n)}_r}},\label{eq:numinv}
\end{align} where the subscripts $p$ and $r$ correspond to the different bins of the distribution. Here the measure $\tilde\rho^{(0)}_r(\tilde k^{(0)}_r)d\tilde k^{(0)}_r$ doesn't need to be updated as it is an invariant. We thus fix it at the initial assumption made for the hidden degrees, taken to have the same values and duplicities as the observed degrees. The order parameters, that are continuous functions of the degree or to be precise, discrete functions of each value of the degree possible, are also calculated simply using these binnings, writing
\begin{align}
    \begin{aligned}
        cB_p&=\sum_r dk_r\rho(k_r)P(\tilde k(k_p)\tilde k(k_r))\langle S_p\rangle,\\
        cq_p&=\sum_r dk_r\rho(k_r)P(\tilde k(k_p)\tilde k(k_r))\langle S_p^2\rangle.
    \end{aligned}
\end{align}
The linear stability of the single equilibrium solution is found to be given by the maximum eigenvalues of the matrix whose terms are given similarly using the bins,
\begin{align}
    \begin{aligned}
        M_{pr}=\frac{\sigma^2}{c}\rho_p\phi_rP_{pr}dk_r.
    \end{aligned}
\end{align}
If this matrix has any eigenvalues greater than 1, the unique fixed point is linearly unstable. We calculate the boundary for the onset of unbounded growth by checking for the divergence of the average species abundances $B_k$.

\end{document}